\documentclass{80SA}
\usepackage{txfonts}

\title{Superconductivity Reinforced by Magnetic Field and the Magnetic Instability in Uranium Ferromagnets}

\author{%
Dai~\textsc{Aoki}$^1$\thanks{E-mail address: dai.aoki@cea.fr}, %
Tatsuma~D.~\textsc{Matsuda}$^{1,2}$,
Fr\'{e}d\'{e}ric~\textsc{Hardy}$^3$,
Christoph~\textsc{Meingast}$^3$,
Valentin~\textsc{Taufour}$^1$,
Elena~\textsc{Hassinger}$^1$,
Ilya~\textsc{Sheikin}$^4$,
Carley~\textsc{Paulsen}$^5$,
Georg~\textsc{Knebel}$^1$,
Hisashi~\textsc{Kotegawa}$^1$, and
Jacques~\textsc{Flouquet}$^1$
}

\inst{%
$^1$INAC/SPSMS, CEA-Grenoble, 17 rue des Martyrs, 38054 Grenoble, France\\
$^2$Advanced Science Research Center, Japan Atomic Energy Agency, Tokai, Ibaraki 319-1195, Japan\\
$^3$KIT-Karlsruhe, Institut f\"{u}r Festk\"{o}rperphysik, 76021 Karlsruhe, Germany\\
$^4$LNCMI-G, CNRS, 25 Rue des Martyrs, 38042 Grenoble, France\\
$^5$Institut N\'{e}el, CNRS, 25 Rue des Martyrs, 38042 Grenoble, France\\
}
\abst{
We review our recent results on ferromagnetic superconductors, URhGe and UCoGe.
High quality single crystals of both compounds were successfully grown.
The specific heat shows a clear jump related to the superconducting transition in UCoGe.
The finite values of $C/T$ at $0\,{\rm K}$ are discussed in terms of the self-induced vortex state and the value of the ordered moment.
With increasing fields for $H\parallel b$-axis in URhGe, the jump of thermal expansion increases and shifts to lower temperature.
The re-entrant and S-shaped superconducting phases for URhGe and UCoGe respectively are explained by the
unusual field dependence of the effective mass, which is induced by 
the ferromagnetic instability when the field is applied along the hard magnetization $b$-axis.
The magnetic fluctuations are very sensitive to the field orientation.
This is reflected in the $H_{\rm c2}$ and the anisotropy of the effective mass.
}

\kword{ferromagnetism, superconductivity, URhGe, UCoGe, effective mass, ferromagnetic instability}

\begin{document}
\maketitle

\section{Introduction}
Ferromagnetism had been thought to be antagonistic to superconductivity, 
because the large internal field destroys the superconducting Cooper pairs easily.
More than three decades ago, 
several ferromagnets, such as ErRh$_4$B$_4$~\cite{Fer77}, Chevrel phase compound HoMo$_6$S$_8$~\cite{Lyn78}, 
were already reported to show superconductivity, however,
the Curie temperature $T_{\rm Curie}$ is lower than 
the superconducting transition temperature $T_{\rm sc}$.
When the ferromagnetic state is well established, the superconducting state is 
expelled, meaning that both phenomena are competitive each other.

UGe$_2$ is the first material which shows coexistence of ferromagnetism and superconductivity.~\cite{Sax00}
$T_{\rm Curie}$ is suppressed with pressure and changes into the first order at tricritical point,
and finally becomes zero at $P_{\rm c}$.
The superconductivity appears below $P_{\rm c}$
displaying that the superconducting state exists only in the ferromagnetic state.
The exotic superconducting state based on the spin-triplet state, 
which is free from the pair-breaking due to the ferromagnetic large internal field, is expected.
Recently we have clarified the existence of the firtst order transition under magnetic fields,
indicating the ``Wing''-shaped $T,P,H$-phase diagram.~\cite{Tau10} (see also ref.~\citen{Kab10})

URhGe with the TiNiSi-type orthorhombic structure also 
shows coexistence of ferromagnetism ($T_{\rm Curie}=9.5\,{\rm K}$) and superconductivity 
($T_{\rm sc}=0.25\,{\rm K}$) at ambient pressure.~\cite{Aok01}
With increasing pressure, $T_{\rm Curie}$ increases while $T_{\rm sc}$ decreases and becomes zero around $4\,{\rm GPa}$,
indicating that URhGe goes away from quantum criticality with pressure.~\cite{Har05_pressure,Miy09}
Instead, the magnetic field for a given crystallographic direction 
is a tuning parameter for quantum criticality at ambient pressure.
When the field is applied along the hard magnetization $b$-axis,
reentrant superconductivity is observed between $8$ and $13\,{\rm T}$,
which is associated with a spin-reorientation at $H_{\rm R}=12\,{\rm T}$.~\cite{Lev05}
Recently, we found an enhancement of the effective mass around $H_{\rm R}$.~\cite{Miy08}
This mass enhancement increases $T_{\rm sc}$ under magnetic field, as a result, superconductivity is reinforced at high fields.
The key points are: i) $H_{\rm c2}$ is governed only by the orbital limit $H_{\rm orb}$,
which is related to the effective mass $m^\ast$ and $T_{\rm sc}$, namely $H_{\rm orb}\sim (m^\ast T_{\rm sc})^2$,
while there is no Pauli paramagnetic limit caused by the Zeeman spin splitting effect,
because the spin-triplet state with the strong Ising character of ferromagnetism is realized;~\cite{Min10}
ii) the magnetic fluctuations associated with the ferromagnetic instability 
develops at high fields around $H_{\rm R}$, and the effective mass is enhanced.

UCoGe is a family of URhGe with the identical crystal structure.
$T_{\rm Curie}$ ($\sim 2.6\,{\rm K}$) and the ordered moment ($\sim 0.05\,\mu_{\rm B}$)
are relatively small, indicating a weak itinerant ferromagnet~\cite{Huy07}.
The superconductivity appears below $T_{\rm sc}\sim 0.6\,{\rm K}$.
The Co-NQR experiments confirm the microscopic coexistence of ferromagnetism and superconductivity.~\cite{Oht08,Oht10}
It is also suggested the ferromagnetic transition is of first order, from the abrupt change of the NQR frequency below $T_{\rm Curie}$.
Contrary to the phase diagram of UGe$_2$ and the previous theoretical prediction~\cite{Fay80},
$T_{\rm sc}$ reveals a broad maximum at $P_{\rm c}$,
where $T_{\rm Curie}$ is suppressed by pressure,
and the superconductivity survives not only in the ferromagnetic state, but also in the paramagnetic state.~\cite{Has08_UCoGe,Slo09,Has10}
A new theory explains this phase diagram from the group symmetry approach.~\cite{Min08}
According to this theory, the first order transition between ferromagnetic-superconducting phase 
and paramagnetic-superconducting phase is inevitably accompanied.
Recently we found very huge and highly anisotropic $H_{\rm c2}$ in UCoGe.~\cite{Aok09_UCoGe}
$H_{\rm c2}$ for $H\parallel c$-axis (easy-magnetization axis) is relatively small, $0.6\,{\rm T}$.
However, when the field is applied along the hard magnetization $a$ and $b$-axis,
$H_{\rm c2}$ reaches $\sim 30\,{\rm T}$ and $\sim 20\,{\rm T}$ at $0\,{\rm K}$, respectively.
In particular, for $H\parallel b$-axis, the $H_{\rm c2}$ is strongly enhanced at $T/T_{\rm sc}\sim 0.4$
with an inverse S-shaped curve.
The resistivity $A$ coefficient, which is proportional to square of effective mass from Kadowaki-Woods ration,
is quite anisotropic.
While the $A$ coefficient for $H\parallel c$-axis is suppressed under magnetic fields, as usually observed
in weak itinerant ferromagnets,
The $A$ coefficient for $H \parallel b$-axis remains high value and shows a maximum around $14\,{\rm T}$
where the S-shaped $H_{\rm c2}$ curve appears.
The reduced $T_{\rm Curie}$ under magnetic fields is connected to the S-shaped $H_{\rm c2}$ curve,
thus it is naively believed that
magnetic fluctuations tuned by the magnetic field reinforces the superconductivity,
which resembles to the reentrant superconductivity in URhGe.

\begin{figure}[tbh]
\begin{center}
\includegraphics[width=1\hsize,clip]{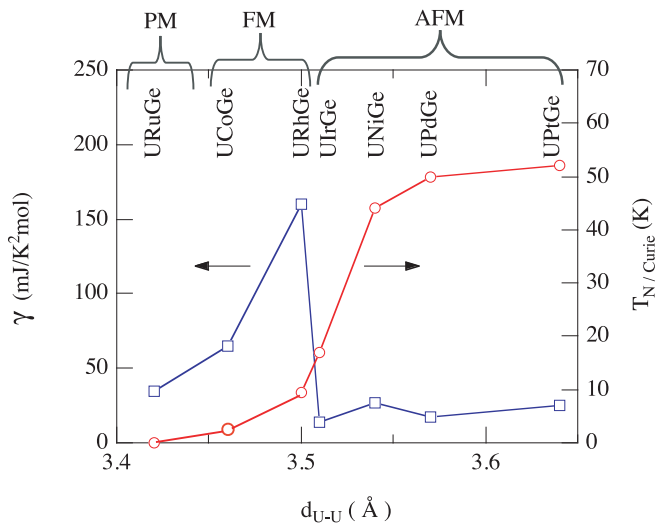}
\end{center}
\caption{(Color online) Sommerfeld coefficient $\gamma$ and the magnetic transition temperature ($T_{\rm N}$ or $T_{\rm Curie}$)
as a function of the distance between the next nearest neighbors of uranium atoms $d_{\rm U-U}$
in UTGe (T: transition metal).  
PM, FM and AM denote paramagnetism, ferromagnetism and antiferromagnetism, respectively.
}
\label{fig:UTGe}
\end{figure}

In UTGe (T: transition metal), the Sommerfeld coefficient $\gamma$
and the magnetic transition temperature (N\'{e}el temperature $T_{\rm N}$ or $T_{\rm Curie}$)
systematically change
as a function of the distance between the next nearest neighbors of uranium atoms $d_{\rm U-U}$,
as shown in Fig.~\ref{fig:UTGe}.
The $d_{\rm U-U}$ of UCoGe and URhGe is close to the so-called Hill limit ($\approx 3.5\,{\rm \AA}$).
Both are located between the paramagnetic state and antiferromagnetic state,
and the ordered temperatures are relatively small.
On the other hand,
the $\gamma$-value in UCoGe and URhGe are relatively large with a maximum value for URhGe, $160\,{\rm mJ/K^2 mol}$,
indicating the heavy electronic state,
which might be favorable for heavy fermion superconductivity.
For UCoGe, the $\gamma$-value is only $60\,{\rm mJ/K^2 mol}$, however,
it is relatively large since the carrier number is small compared to URhGe.~\cite{Aok10_UCoGe,Sam10}

\section{Experimental}
\begin{figure}[tbh]
\begin{center}
\includegraphics[width=1 \hsize,clip]{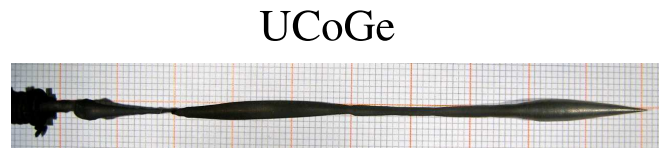}
\end{center}
\caption{(Color online) Single crystal of UCoGe.}
\label{fig:UCoGe}
\end{figure}

High-quality single crystals of URhGe and UCoGe were grown using the Czochralski method in a tetra-arc furnace. 
The starting materials with appropriate ratio were melted under purified Ar atmosphere gas.
The polycrystalline ingot was turned over and melted again. 
This process was repeated several times in order to obtain a homogeneous phase.
Then the single crystal was pulled using a seed crystal at a slow pulling rate $~15\,{\rm mm/hr}$.
The obtained single crystal of UCoGe is shown in Fig.~\ref{fig:UCoGe}.
Single crystals were subsequently annealed under ultra high vacuum (UHV) in a horizontal radio frequency furnace at a temperature just below the melting point for 12 hrs.
Single crystals were then annealed again under UHV for 20 days at $900\,^\circ{\rm C}$ in an electrical furnace.
Single crystals were cut by a spark cutter.
Since both URhGe and UCoGe are most likely incongruent melt,
it is very difficult to obtain high-quality single crystals.
We cut the single crystals into many small pieces,
and looked for the high quality parts by resistivity and Laue photograph.
The residual resistivity ratios (RRR) were $40$-$65$ in URhGe and $30$ in UCoGe.
Resistivity measurements were done by four-probe AC method at low temperatures down to $70\,{\rm mK}$ and at high fields up to $16\,{\rm T}$.
Specific heat measurements were done using relaxation and semi-adiabatic methods down to $100\,{\rm mK}$.
Thermal expansion measurements in URhGe were done using strain gauges with the active dummy method.
Under magnetic field, the field direction was carefully controlled by a rotation mechanism with the precision of $0.2\,{\rm deg}$ in order to get perfectly alignment with the $b$-axis.

\section{Results and Discussion}
\begin{figure}[tbh]
\begin{center}
\includegraphics[width=1 \hsize,clip]{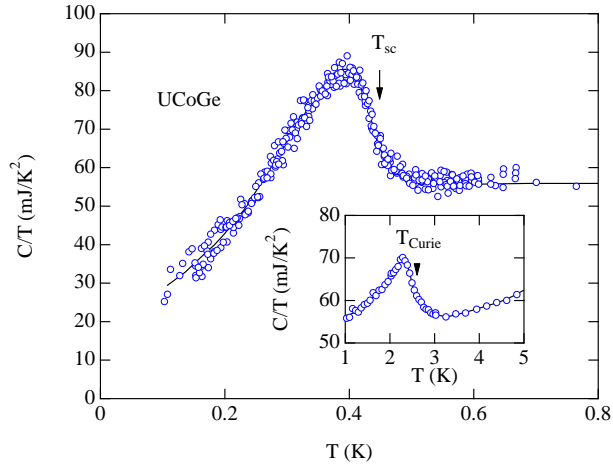}
\end{center}
\caption{(Color online) Temperature dependence of the specific heat in UCoGe in the form of $C/T$ vs $T$.}
\label{fig:UCoGe_Cp}
\end{figure}

Figure~\ref{fig:UCoGe_Cp} shows the temperature dependence of the specific heat for UCoGe.
At high temperature, the specific heat shows the small jump at $2.6\,{\rm K}$,
related to the Curie temperature $T_{\rm Curie}$.
The magnetic entropy is also small, indicating a typical weak ferromagnetic behavior.
At low temperature, the specific heat shows the jump again due to the superconducting transition
at $T_{\rm sc}=0.44\,{\rm K}$.
The value obtained in specific heat measurement corresponding to the bulk properties
is lower than $T_{\rm sc}$ of resistivity,
which is typically $0.6\,{\rm K}$.
This discrepancy is usually observed in many heavy fermion superconductors, such as URu$_2$Si$_2$.
The value of $\Delta C/\gamma T_{\rm sc}$ ($\sim 0.69$) is much smaller than
the value expected from the weak-coupling BCS model ($1.43$).
It is worth noting that the small value ($\sim 0.6$) is also observed in URhGe.
The finite value of $C/T$ at $0\,{\rm K}$ is approximately $20$-$30\,{\rm mJ/K^2 mol}$.
The important question is whether this is intrinsic or not.
Of course, it is known that the finite value of $C/T$ strongly depends on the sample quality.
However, if we believe that the finite value is intrinsic, 
two possible origins can be considered:
i) one of the spin Fermi surface is gapped, and the other is not gapped. 
In ferromagnetic superconductors, the spin triplet state is naturally believed, 
and the equal spin pairing should be responsible for the superconductivity, because of the absence of Pauli limit.
If the spin splitting is large due to the ferromagnetism, only one spin might be responsible for superconductivity and the other might be not, 
namely $\Delta_{\downarrow\downarrow} \neq 0$ and $\Delta_{\uparrow\uparrow}\approx 0$.
ii) another origin is due to the self-induced vortex state,
where there is no  $H_{\rm c1}$ and the material is always in the superconducting mixed state. 
In fact, the self-induced vortex state is suggested in UCoGe by NQR experiments~\cite{Oht10}.
According to the recent magnetization measurement, it is also clear that there is no $H_{\rm c1}$~\cite{Deg10,Car_pub}.

\begin{figure}[tbh]
\begin{center}
\includegraphics[width=1 \hsize,clip]{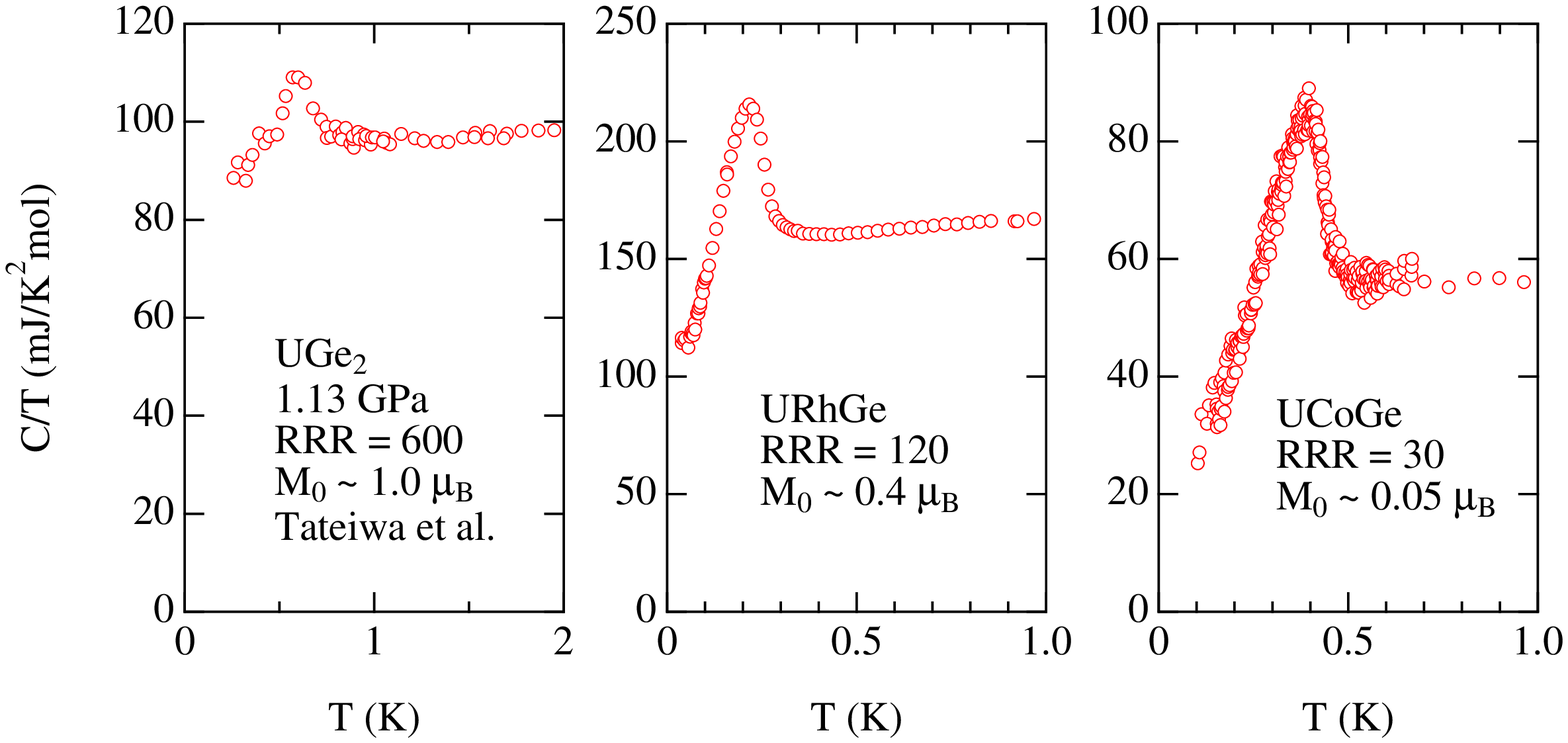}
\end{center}
\caption{(Color online) Temperature dependence of the specific heat at low temperatures in UGe$_2$, URhGe and UCoGe. The data of UGe$_2$ were cited from ref.\citen{Tat01}.}
\label{fig:Cp_all}
\end{figure}

In this sense, it is interesting to compare the specific heat results
for three ferromagnetic superconductors, UGe$_2$, URhGe and UCoGe,
which are shown in Fig.~\ref{fig:Cp_all}.
The data of UGe$_2$ cited from ref.~\citen{Tat01} are results under pressure,
where the ordered moment is approximately $1\,\mu_{\rm B}$~\cite{Pfl02}.
Although the sample is of high quality (${\rm RRR}=600$)
with large mean free path ($l > 1000\,{\rm \AA}$),
the finite value of $C/T$ is relatively large, 
approximately $70\,{\%}$ of density of state is remained at $0\,{\rm K}$.
On the other hand, URhGe shows a smaller residual value, which corresponds to $60\,{\%}$ of the normal state. The ordered moment of URhGe is $0.4\,\mu_{\rm B}$.
Furthermore, in UCoGe, the residual value is approximately $40\,{\%}$, 
which is even smaller than those of UGe$_2$ and URhGe,
although the quality of the sample is less than those of UGe$_2$ and URhGe.
The ordered moment of UCoGe is $0.05\,\mu_{\rm B}$.
Although the sample quality must be improved at least in URhGe and UCoGe,
it seems there is a link between the finite value of $C/T$ and the ordered moment.
That is, when the ordered moment is large, the finite value of $C/T$ is also large.
This relation is consistent with the scenarios based on the self-induced vortex state
or non-gapped majority Fermi surface.

\begin{figure}[tbh]
\begin{center}
\includegraphics[width=0.9 \hsize,clip]{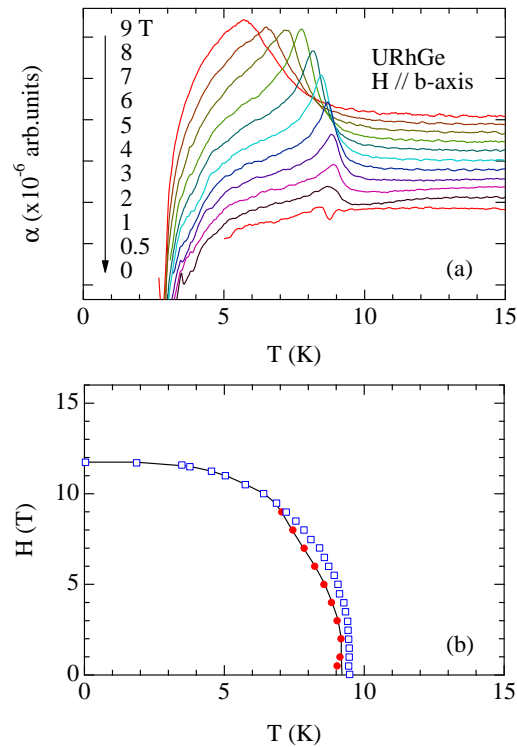}
\end{center}
\caption{(Color online) (a)Thermal expansion measurements for $H\parallel b$-axis in URhGe. (b)Decrease of the Curie temperature under magnetic fields. Solid circles and open squares are from the thermal expansion measurements and magnetization measurements\protect\cite{Har_pub}, respectively.}
\label{fig:URhGe_Texp}
\end{figure}

Let us focus on the high temperature properties of URhGe.
Figure~\ref{fig:URhGe_Texp}(a) shows the temperature dependence of the thermal expansion coefficient $\alpha$ for the field parallel to the hard-magnetization axis ($b$-axis).
At low fields, $\alpha$ shows a relatively small jump, which corresponds to the Curie temperature $T_{\rm Curie}$.
With increasing fields, the jump becomes larger and is shifted to lower temperatures.
The field dependence of $T_{\rm Curie}$ is shown in Fig.\ref{fig:URhGe_Texp}(b).
At low temperatures, the decreased $T_{\rm Curie}$ under magnetic fields is linked to the spin reorientation field $H_{\rm R}\approx 12\,{\rm T}$.
A recent theory based on the Landau free energy analysis explains that
$T_{\rm Curie}$ is suppressed by the field with $T_{\rm Curie}(H)=T_{\rm Curie}(0)-\alpha H^2$,
where $\alpha$ is a constant\cite{Min10_arXiv},
when the field direction is perfectly aligned with the hard magnetization axis ($b$-axis).
The large jump under magnetic fields indicate the volume change at $T_{\rm Curie}$ is large, 
compared to that at zero field,
suggesting that the 2nd order transition at zero field may change into the 1st order transition under magnetic fields.
However, no discontinuity in the magnetization can be found in the extrapolation to $0\,{\rm K}$~\cite{Har_pub}.

In UGe$_2$, it is known that $T_{\rm Curie}$ decreases with pressure 
and changes from 2nd order to 1st order at the tricritical point.
Under magnetic fields, the 1st order plane in $T,P,H$ phase diagram emerges and disappears 
at quantum critical point.~\cite{Bel05,Tau10}
By analogy with UGe$_2$, a similar phase diagram is suggested in URhGe~\cite{Lev07}.

\begin{figure}[tbh]
\begin{center}
\includegraphics[width=0.9 \hsize,clip]{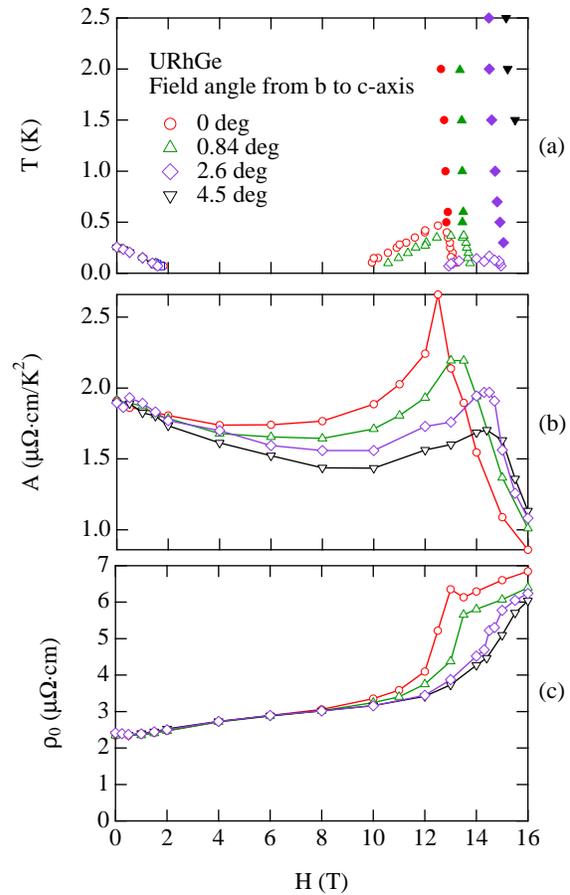}
\end{center}
\caption{(Color online) (a)Field-temperature phase diagram of URhGe for the field direction close to the $b$-axis. The corresponding field dependence of resistivity $A$ coefficient (b) and the residual resistivity $\rho_0$ (c)}
\label{fig:URhGe_A_angdep}
\end{figure}

Figure~\ref{fig:URhGe_A_angdep}(a) shows the phase diagram of URhGe at high fields and at low temperatures obtained from resistivity measurements.
When the field is almost perfectly aligned to $b$-axis,
the re-entrant superconductivity emerges between $8$ to $13\,{\rm T}$.
The maximum temperature of the re-entrant superconducting phase is $T_{\rm rsc,max}=0.45\,{\rm K}$,
which is larger than the low field value $T_{\rm sc}=0.25\,{\rm K}$.
The spin re-orientation field is $H_{\rm R}=13\,{\rm T}$.
With tilting the field direction from $b$ to $c$-axis,
the re-entrant superconducting phase shifts to higher field and shrinks.
At $4.5\,{\rm deg}$, the re-entrant superconducting phase completely disappears.
$T_{\rm rsc,max}$ linearly decreases with the field angle and is expected to be zero at $4\,{\rm deg}$.
Correspondingly, $H_{\rm R}$ also shifts to the higher fields.

In a previous report~\cite{Miy08}, we have clarified that the mass enhancement plays an important role
to induce re-entrant superconductivity. 
On the basis of the McMillan-like formula, $T_{\rm sc}$ is expected to increase
when the electron correlation is enhanced.
We can simply describe $T_{\rm sc}$ as $T_{\rm sc}\propto \exp(-m^\ast/m^{\ast\ast})$,
where the effective mass $m^\ast$ and the correlation mass $m^{\ast\ast}$ 
are related with the band mass $m_{\rm b}$ as $m^\ast=m_{\rm b}+m^{\ast\ast}$.
In ferromagnetic superconductors in the spin-triplet state, 
$H_{\rm c2}$ is simply limited by the orbital limit,
thus we can describe as $H_{\rm c2}\sim (m^\ast T_{\rm sc})^2$.
Therefore, the enhancement of effective mass increases $H_{\rm c2}$.
It must be noted that we assume that the Fermi surface and the band mass are unchanged.
Figure~\ref{fig:URhGe_A_angdep}(b) shows the field dependence of the $A$ coefficient from the resistivity measurements,
which always show the Fermi liquid behavior at least below $1\,{\rm K}$ within the measured field angle.
According to the Kadowaki-Woods ratio, the $A$ coefficient has a relation with the $\gamma$-value, namely $A\propto \gamma^2 \propto \sqrt{m^\ast}$.

At $0\,{\rm deg}$, the $A$ coefficient reveals a clear maximum at $H_{A,{\rm max}}=12.5\,{\rm T}$
and is strongly suppressed by further increasing field.
The re-entrant superconductivity emerges when the $A$ coefficient exceeds a critical value.
By rotating the field angle, $H_{A,{\rm max}}$ shifts to higher fields,
and the peak structure of the field variation of $A$ is smeared out.
This behavior also corresponds to the re-entrant superconducting phase,
which shifts to higher fields and shrinks with increasing the tilted angle.

We have reported a similar behavior in the resistivity measurements under pressure.~\cite{Miy09}
Applying pressure, the reentrant superconducting phase shrinks and shifts to the higher field,
and finally disappears at $1.8\,{\rm GPa}$.
$H_{A,{\rm max}}$ also shifts to the higher field and the maximum of $A$ is reduced.
At zero field $T_{\rm Curie}$ increases and the low field superconducting $T_{\rm sc}$ decreases with increasing pressure,
meaning that URhGe moves away from a critical point with pressure.
The present results indicate that the field angle from $b$ to $c$-axis also plays a similar role,
as pressure.
The small deviation from $H\parallel b$-axis to $c$-axis strongly suppresses 
the magnetic instability and the mass enhancement.

\begin{figure}[tbh]
\begin{center}
\includegraphics[width=0.9 \hsize,clip]{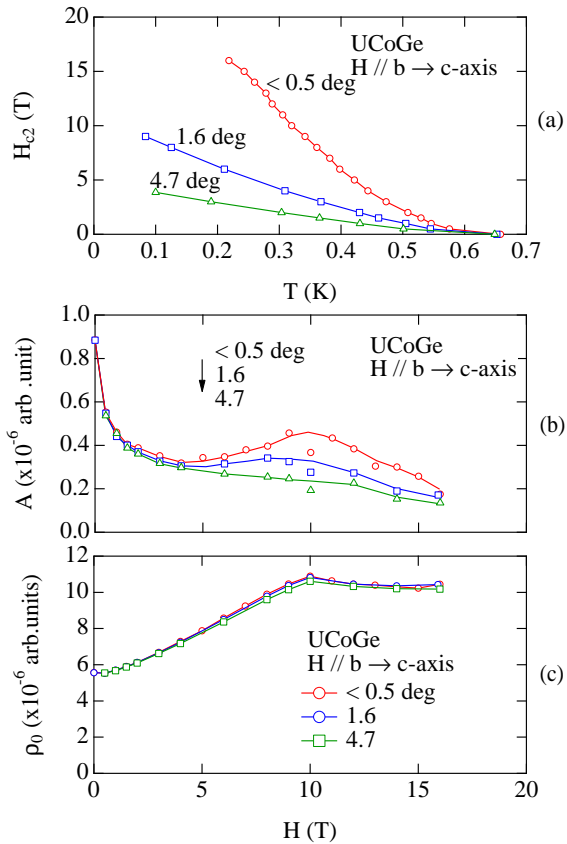}
\end{center}
\caption{(Color online) (a)Temperature dependence of $H_{\rm c2}$ for different field angles from $b$ to $c$-axis in UCoGe. (b)Field dependence of $A$ coefficient and (c)field dependence of the residual resistivity $\rho_0$.}
\label{fig:UCoGe_A_angdep}
\end{figure}

In UCoGe, the small deviation from $b$ to $c$-axis affects the $H_{\rm c2}$ curve more drastically.
As shown in Fig.~\ref{fig:UCoGe_A_angdep}(a),
$H_{\rm c2}$ for field angle less than $0.5\,{\rm deg}$
shows an S-shaped curve and reaches $H_{\rm c2}(0)\approx 20\,{\rm T}$.
In our previous report with a different sample,~\cite{Aok09_UCoGe} 
the S-shaped curve is more significant than the present result.
This is most likely due to the sample quality and the optimization of the field angle.
By tilting the field angle slightly, $H_{\rm c2}$ is strongly suppressed,
correspondingly the peak structure of the $A$ coefficient is suppressed,
as shown in Fig.~\ref{fig:UCoGe_A_angdep}(b).
The huge $H_{\rm c2}$ is basically understood by the enhancement of effective mass under magnetic fields as we already mentioned in URhGe.
Compared to URhGe,
the specific points in UCoGe are: i) UCoGe is already close to the quantum critical point,~\cite{Has08_UCoGe} 
ii) UCoGe is most likely a low carrier system,~\cite{Sam10,Aok10_UCoGe} thus the field dependence of band mass is probably more significant than
the case of URhGe.

\begin{figure}[tbh]
\begin{center}
\includegraphics[width=1 \hsize,clip]{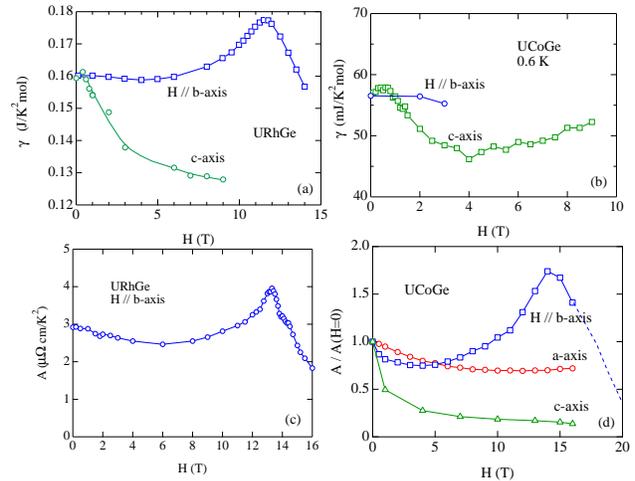}
\end{center}
\caption{(Color online) Field dependence of the $\gamma$-values for $H \parallel b$ and $c$-axis in URhGe(a) and in UCoGe(b). The data for $H\parallel b$-axis in URhGe was obtained from the magnetization measurement using the thermodynamic Maxwell relation~\protect\cite{Har_pub}, assuming $\gamma =160\,{\rm mJ/K^2 mol}$ at zero field.
The field dependence of the $A$ coefficient in URhGe(c) and UCoGe(d)~\protect\cite{Aok09_UCoGe}.}
\label{fig:gamma_A_Hdep}
\end{figure}

In order to compare the field response of the effective mass for different field direction, namely
for hard-magnetization $b$-axis and easy-magnetization $c$-axis,
we show in Fig.~\ref{fig:gamma_A_Hdep} the field dependence of the $\gamma$-value
and the $A$ coefficients for $H \parallel b$ and $c$-axis in URhGe and UCoGe.
The $\gamma$-value for $H\parallel b$-axis in URhGe 
is obtained from the temperature dependence of magnetization.~\cite{Har_pub}
From the thermodynamic Maxwell relation~\cite{Pau90}, $\partial M/\partial T = \partial S/ \partial B$,
we can describe the $\gamma$-value, as $\partial \gamma/\partial B = 2\beta$,
where $\beta$ is a coefficient of $T^2$ term of magnetization, namely
$M=M(0)+\beta T^2$.
In URhGe, it it clear that the effective mass shows a maximum at $H_{\rm R}\approx 12\,{\rm T}$
for $H\parallel b$-axis both from the $\gamma$-value and from the $A$ coefficient,
indicating that magnetic fluctuations are induced by the magnetic field
For $H\parallel c$-axis, the $\gamma$-value decreases monotonously down to $130\,{\rm mJ/K^2 mol}$ at $9\,{\rm T}$.
This is usually observed in weak ferromagnets.

In UCoGe, the field dependence of $\gamma$-value looks similar to that of URhGe,
although the data for $H\parallel b$-axis is only up to $3\,{\rm T}$.
The $\gamma$-value for $H\parallel b$ remains relatively large,
but for $H\parallel c$-axis, it decreases at least down to $4\,{\rm T}$.
Above $4\,{\rm T}$, the value starts to increase slightly, which is not observed in
the field variation of the $A$ coefficient.
Polarized neutron scattering experiments for $H\parallel c$-axis shows that
the magnetic moment on the Co-site is induced in antiparallel direction at $12\,{\rm T}$~\cite{Pro10}.
The magnetic properties in UCoGe are more complicated than that in URhGe.
This may be also related with the small carrier number in UCoGe.
The small increase of $\gamma$-value above $4\,{\rm T}$ can be 
due to such a complicated magnetic properties.

\begin{figure}[tbh]
\begin{center}
\includegraphics[width=1 \hsize,clip]{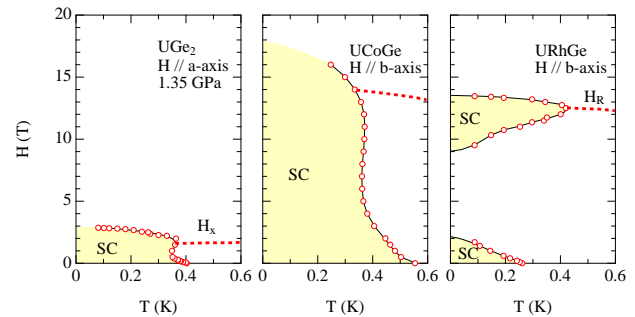}
\end{center}
\caption{(Color online) Field-temperature phase diagrams of UGe$_2$~\protect\cite{She01}, UCoGe~\protect\cite{Aok09_UCoGe} and URhGe. SC denotes the superconducting phase, and the dotted lines are the magnetic anomalies.
In UGe$_2$, the field is applied to the easy-magnetization axis ($a$-axis),
and the pressure is between $P_{\rm x}$ and $P_{\rm c}$. 
With increasing pressure, the ferromagnetic phase changes from FM1 (weakly polarized) phase to FM2 (strongly polarized) phase.}
\label{fig:UGe2_UCoGe_URhGe_SC_phase}
\end{figure}

Finally, we show in Fig.~\ref{fig:UGe2_UCoGe_URhGe_SC_phase} the superconducting phase diagram 
of three ferromagnetic superconductors, UGe$_2$, UCoGe and URhGe.
In UGe$_2$, the field is applied along the easy-magnetization $a$-axis and 
the pressure is between $P_{\rm x}$ and $P_{\rm c}$.
With increasing field, it is known that the ferromagnetic phase 
changes from FM1 (weakly polarized) phase to FM2 (strongly polarized phase).
The dotted line corresponds to the so-called $H_{\rm x}$.
The $H_{\rm c2}$ curve is also unusual, indicating an abrupt enhancement with
an S-shaped curve.
The $A$ coefficient displays a maximum at $H_{\rm x}$ 
and then is immediately suppressed above $H_{\rm x}$.~\cite{Ter06}
This is similar to the data on URhGe and UCoGe.
However, there is a drastic change of Fermi surface between FM1 and FM2 phase~\cite{Ter02,Set02},
meaning that both $k_{\rm F}$ and the band mass also change below and above $H_{\rm x}$.
The response to $H_{\rm c2}$ curve might be more complicated than those of URhGe and UCoGe.
On the other hand, in URhGe and UCoGe, we can simply explain $H_{\rm c2}$ curve from the field dependence of the effective mass,
where we assume unchanged Fermi surfaces and the constant band mass with field.
Up to now, there is no direct evidence that the Fermi surface is unchanged. 
In particular, if $H_{\rm R}$ in URhGe is not a first order transition,
the preservation of Fermi surface is a good approximation.

\section{Summary}
We have shown our recent results in ferromagnetic superconductors, focusing on UCoGe and URhGe.
Unusual behaviors of $H_{\rm c2}$ curve is explained by the field dependence of the effective mass.
The spin triplet state with equal-spin pairing is most likely realized,
since $H_{\rm c2}$ is governed only by the orbital limit in the absence of Pauli limitation.
When the field is applied along the hard magnetization axis, 
magnetic fluctuations are induced and thus the effective mass shows the maximum at a certain field.
This is very sensitive to the field orientation and the slight tilt to the easy magnetization axis strongly
suppresses both magnetic fluctuations and superconducting phase.
It is worth noting that a recent theory explains the large anisotropy of $H_{\rm c2}$
on the basis of a strong coupling behavior for $H\parallel a$-axis near ferromagnetic quantum critical points.~\cite{Tad_ICHE}

\section*{Acknowledgements}
We thank J. P. Brison, L. Howald, L. Malone, W. Knafo, K. Ishida, Y. Tada, S. Fujimoto and H. Harima for helpful discussion.
This work was supported by ERC starting grant (NewHeavyFermion) and French ANR project (CORMAT, SINUS).


\begin{thebibliography}{10}
\bibitem{Fer77}
W.~A. Fertig, D.~C. Johnston, L.~E. DeLong, R.~W. McCallum, M.~B. Maple and
  B.~T. Matthias: Phys. Rev. Lett. {\bf 38} (1977) 987.

\bibitem{Lyn78}
J.~W. Lynn, D.~E. Moncton, W.~Thomlinson, G.~Shirane and R.~N. Shelton: Solid
  State Commun. {\bf 26}~(8) (1978) 493.

\bibitem{Sax00}
S.~S. Saxena, P.~Agarwal, K.~Ahilan, F.~M. Grosche, R.~K.~W. Haselwimmer, M.~J.
  Steiner, E.~Pugh, I.~R. Walker, S.~R. Julian, P.~Monthoux, G.~G. Lonzarich,
  A.~Huxley, I.~Sheikin, D.~Braithwaite and J.~Flouquet: Nature {\bf 406}
  (2000) 587.

\bibitem{Tau10} 
V. Taufour, D. Aoki, G. Knebel and J. Flouquet:
Phys. Rev. Lett. {\bf 105} (2010) 217201.


\bibitem{Kab10}
N.~Kabeya, R.~Iijima, E.~Osaki, S.~Ban, K.~Imura, K.~Deguchi, N.~Aso, Y.~Homma,
  Y.~Shiokawa and N.~K. Sato: J. Phys.: Conf. Ser. {\bf 200} (2010) 032028.

\bibitem{Aok01}
D.~Aoki, A.~Huxley, E.~Ressouche, D.~Braithwaite, J.~Flouquet, J.-P. Brison,
  E.~Lhotel and C.~Paulsen: Nature {\bf 413} (2001) 613.

\bibitem{Har05_pressure}
F.~Hardy, A.~Huxley, J.~Flouquet, B.~Salce, G.~Knebel, D.~Braithwaite, D.~Aoki,
  M.~Uhlarz and C.~Pfleiderer: Physica B {\bf 359} (2005) 1111.

\bibitem{Miy09}
A.~Miyake, D.~Aoki and J.~Flouquet: J. Phys. Soc. Jpn. {\bf 78} (2009) 063703.

\bibitem{Lev05}
F.~L\'{e}vy, I.~Sheikin, B.~Grenier and A.~D. Huxley: Science {\bf 309} (2005)
  1343.

\bibitem{Miy08}
A.~Miyake, D.~Aoki and J.~Flouquet: J. Phys. Soc. Jpn. {\bf 77} (2008) 094709.

\bibitem{Min10}
V.~P. Mineev: Phys. Rev. B {\bf 81}~(18) (2010) 180504.

\bibitem{Huy07}
N.~T. Huy, A.~Gasparini, D.~E. {de Nijs}, Y.~Huang, J.~C.~P. Klaasse,
  T.~Gortenmulder, A.~{de Visser}, A.~Hamann, T.~{G\"{o}rlach} and
  H.~v.~{L\"{o}hneysen}: Phys. Rev. Lett. {\bf 99} (2007) 067006.

\bibitem{Oht08}
T.~Ohta, Y.~Nakai, Y.~Ihara, K.~Ishida, K.~Deguchi, N.~K. Sato and I.~Satoh: J.
  Phys. Soc. Jpn. {\bf 77} (2008) 023707.

\bibitem{Oht10}
T.~Ohta, T.~Hattori, K.~Ishida, Y.~Nakai, E.~Osaki, K.~Deguchi, N.~K. Sato and
  I.~Satoh: J. Phys. Soc. Jpn. {\bf 79}~(2) (2010) 023707.

\bibitem{Fay80}
D.~Fay and J.~Appel: Phys. Rev. B {\bf 22} (1980) 3173.

\bibitem{Has08_UCoGe}
E.~Hassinger, D.~Aoki, G.~Knebel and J.~Flouquet: J. Phys. Soc. Jpn. {\bf
  77}~({7}) (2008) 073703.

\bibitem{Slo09}
E.~Slooten, T.~Naka, A.~Gasparini, Y.~K. Huang and A.~de~Visser: Phys. Rev.
  Lett. {\bf 103}~(9) (2009) 097003.

\bibitem{Has10}
E.~Hassinger, D.~Aoki, G.~Knebel and J.~Flouquet: J. Phys.: Conf. Ser. {\bf
  200} (2010) 012055.

\bibitem{Min08}
V.~P. Mineev: J. Phys. Soc. Jpn. {\bf 77} (2008) 103702.

\bibitem{Aok09_UCoGe}
D.~Aoki, T.~D. Matsuda, V.~Taufour, E.~Hassinger, G.~Knebel and J.~Flouquet: J.
  Phys. Soc. Jpn. {\bf 78} (2009) 113709.

\bibitem{Aok10_UCoGe}
D. Aoki, I. Sheikin, T. D. Matsuda, V. Taufour, G. Knebel and J. Flouquet:
to be published in J .Phys. Soc. Jpn. No.1 (2011).

\bibitem{Sam10}
M.~{Samsel-Czeka{\l}a}, S.~Elgazzar, P.~M. Oppeneer, E.~Talik, W.~Walerczyk and
  R.~Tro{\'c}: J. Phys.: Condens. Matter {\bf 22} (2010) 015503.

\bibitem{Deg10} 
K. Deguchi, E. Osaki, S. Ban, N. Tamura, Y. Simura, T. Sakakibara, I. Satoh and N. K. Sato:
J. Phys. Soc. Jpn. {\bf 79} (2010) 083708.

\bibitem{Car_pub}
C. Paulsen \textit{et al.}: to be published.

\bibitem{Tat01}
N.~Tateiwa, T.~C. Kobayashi, K.~Hanazono, K.~Amaya, Y.~Haga, R.~Settai and
  Y.~\={O}nuki: J. Phys.: Condens. Matter {\bf 13} (2001) L17.

\bibitem{Min10_arXiv}
V.P. Mineev: arXiv:1002.3510v1.

\bibitem{Pfl02}
C.~Pfleiderer and A.~D. Huxley: Phys. Rev. Lett. {\bf 89} (2002) 147005.

\bibitem{Har_pub}
F. Hardy \textit{et al.}: to be published.

\bibitem{Bel05}
D.~Belitz, T.~R. Kirkpatrick and {J\"{o}rg Rollb\"{u}hler}: Phys. Rev. Lett.
  {\bf 94}~(24) (2005) 247205.

\bibitem{Lev07}
F.~L\'{e}vy, I.~Sheikin and A.~Huxley: Nature Physics {\bf 3} (2007) 460.

\bibitem{Pau90}
C.~Paulsen, A.~Lacerda, L.~Puech, P.~Haen, P.~Lejay, J.~L. Tholence,
  J.~Flouquet and A.~{de Visser}: J. Low Temp. Phys. {\bf 81} (1990) 317.

\bibitem{Pro10}
K.~Proke\v{s}, A.~{de Visser}, Y.~K. Huang, B.~F{\aa}k and E.~Ressouche: Phys.
  Rev. B {\bf 81}~(18) (2010) 180407.

\bibitem{Ter06}
T.~Terashima, K.~Enomoto, T.~Konoike, T.~Matsumoto, S.~Uji, N.~Kimura, M.~Endo,
  T.~Komatsubara, H.~Aoki and K.~Maezawa: Phys. Rev. B {\bf 73} (2006) 140406.

\bibitem{Ter02}
T.~Terashima, T.~Matsumoto, C.~Terakura, S.~Uji, N.~Kimura, M.~Endo,
  T.~Komatsubara, H.~Aoki and K.~Maezawa: Phys. Rev. B {\bf 65} (2002) 174501.

\bibitem{Set02}
R.~Settai, M.~Nakashima, S.~Araki, Y.~Haga, T.~C. Kobayashi, N.~Tateiwa,
  H.~Yamagami and Y.~\={O}nuki: J. Phys.: Condens. Matter {\bf 14} (2002) L29.

\bibitem{She01}
I.~Sheikin, A.~Huxley, D.~Braithwaite, J.~P. Brison, S.~Watanabe, K.~Miyake and
  J.~Flouquet: Phys. Rev. B {\bf 64} (2001) 220503.

\bibitem{Tad_ICHE}
Y.~Tada, N.~Kawakami and S.~Fujimoto: arXiv:1008.4204.

\end{thebibliography}

\end{document}